\documentclass[twocolumn]{article}
\usepackage{natbib}

\usepackage{amsmath}
\usepackage{amsfonts}
\usepackage{graphicx}
\usepackage{color}
\usepackage{caption}
\usepackage[caption=false]{subfig}
\usepackage[letterpaper, margin=1in]{geometry}

\usepackage{wrapfig}

\definecolor{darkgreen}{rgb}{0,0.5,0}

\renewcommand{\dim}{\text{\dim}}

\usepackage{authblk}

\begin{document}
\title{Two's company, three (or more) is a simplex: Algebraic-topological tools for understanding higher-order structure in neural data}
\author{Chad Giusti \and Robert Ghrist \and Danielle S. Bassett}

\date{\today}
\twocolumn[
\begin{@twocolumnfalse}
\maketitle

\begin{abstract}

The language of graph theory, or network science, has proven to be an exceptional tool for addressing myriad problems in neuroscience. Yet, the use of networks is predicated on a critical simplifying assumption: that the quintessential unit of interest in a brain is a dyad -- two nodes (neurons or brain regions) connected by an edge. While rarely mentioned, this fundamental assumption inherently limits the types of neural structure and function that graphs can be used to model. Here, we describe a generalization of graphs that overcomes these limitations, thereby offering a broad range of new possibilities in terms of modeling and measuring neural phenomena. Specifically, we explore the use of \emph{simplicial complexes}, a theoretical notion developed in the field of mathematics known as algebraic topology, which is now becoming applicable to real data due to a rapidly growing computational toolset. We review the underlying mathematical formalism as well as the budding literature applying simplicial complexes to neural data, from electrophysiological recordings in animal models to hemodynamic fluctuations in humans. Based on the exceptional flexibility of the tools and recent ground-breaking insights into neural function, we posit that this framework has the potential to eclipse graph theory in unraveling the fundamental mysteries of cognition.

\bigskip
\end{abstract}
\end{@twocolumnfalse}
]

\begin{figure*}
\begin{center}
\includegraphics{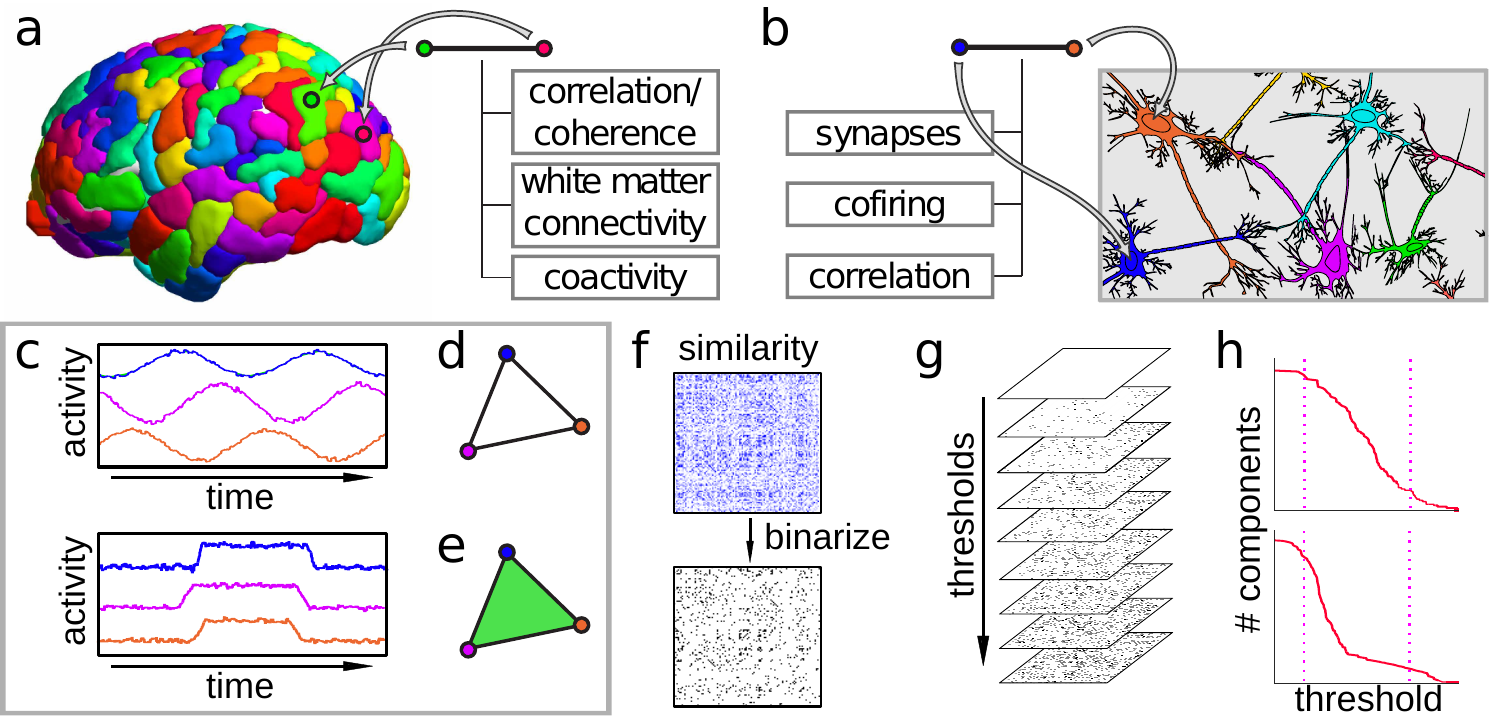}
\end{center}
\caption{Extensions of network models provide insights into neural data. \emph{(a)} Network models are increasingly common for the study of whole-brain activity. \emph{(b)} Neuron-level networks have been a driving force in the adoption of network techniques in neuroscience.  \emph{(c)} Two potential activity traces for a trio of neural units. (top) Activity for a ``pacemaker''-like circuit, whose elements are pairwise active in all combinations but never as a triple. (bottom) Activity for units driven by a common strong stimulus, thus are simultaneously coactive. \emph{(d)} A network representation of the coactivity patterns for either population in (c). Networks are capable of encoding only dyadic relationships, so do not capture the difference between these two populations. \emph{(e)} A \emph{simplicial complex} model is capable of encoding higher order interactions, thus distinguishing between the top and bottom panels in (c). \emph{(f)} A similarity measure for elements in a large neural population is encoded as a matrix, thought of as the adjacency matrix for a complete, weighted network, and binarized using some threshold to simplify quantitative analysis of the system. In the absence of complete understanding of a system, it is difficult or impossible to make a principled choice of threshold value. \emph{(g)} A \emph{filtration} of networks is obtained by thresholding at every possible  entry and arranging the resulting family of networks along an axis at their threshold values. This structure discards no information from the original weighted network. \emph{(g)} Graphs of the number of connected components as a function of threshold value for two networks reveals differences in their structure: (top) homogeneous network {\it versus} (bottom) a modular network. (dotted lines) Thresholding near these values would suggest inaccurately that these two networks have similar structure.}
\label{F:intro}
\end{figure*}

The recent development of novel imaging techniques and the acquisition of massive collections of neural data make finding new approaches to understanding neural structure a vital undertaking. Network science is rapidly becoming an ubiquitous tool for understanding the structure of complex neural systems. Encoding relationships between objects of interest using graphs (Fig. \ref{F:intro}a-\ref{F:intro}b, Fig. \ref{F:sim_cx}a) enables the use of a bevy of well-developed tools for structural characterization as well as inference of dynamic behavior. Over the last decade, network models have demonstrated broad utility in uncovering fundamental architectural principles \citep{Bassett2006b,Bullmore2011} and their implications for cognition \citep{Medaglia2015} and disease \citep{Stam2014}. Their use has led to the development of novel diagnostic biomarkers \citep{Stam2014} and conceptual cognitive frameworks \citep{Sporns2014} that illustrate a paradigm shift in systems, cognitive, and clinical neuroscience: namely, that brain function and alteration are inherently networked phenomena.

All graph-based models consist of a choice of \emph{vertices}, which represent the objects of study, and a collection of \emph{edges}, which encode the existence of a relationship between pairs of objects (Fig. \ref{F:intro}a-\ref{F:intro}b, Fig. \ref{F:sim_cx}a). However, in many real systems, such \emph{dyadic} relationships fail to accurately capture the rich nature of the system's organization; indeed, even when the underlying structure of a system is known to be dyadic, its function is often understood to be polyadic. In large-scale neuroimaging, for example, cognitive functions appear to be performed by a distributed set of brain regions \citep{Gazzaniga2013b} and their interactions \citep{Medaglia2015}. At a smaller scale, the spatiotemporal patterns of interactions between a few neurons is thought to underlie basic information coding \citep{Szatmary2010} and explain alterations in neural architecture that accompany development \citep{Feldt2011}.

Drawing on techniques from the field of \emph{algebraic topology}, we describe a mathematically well-studied generalization of graphs called \emph{simplicial complexes} as an alternative, often preferred method for encoding non-dyadic relationships (Fig. \ref{F:sim_cx}). Different types of complexes can be used to encode co-firing of neurons \citep{curto2008cell}, co-activation of brain areas \citep{Crossley2013}, and structural and functional connections between neurons or brain regions \citep{Bullmore2009} (Fig. \ref{F:data}). After choosing the complex of interest, quantitative and theoretical tools can be used to describe, compare, and explain the statistical properties of their structure in a manner analogous to graph statistics or network diagnostics.

We then turn our attention to a method of using additional data, such as temporal processes or frequency of observations, to decompose a simplicial complex into constituent pieces, called a \emph{filtration} of the complex (Fig. \ref{F:intro}f-\ref{F:intro}h). Filtrations reveal more detailed structure in the complex, and provide tools for understanding how that structure arises (Fig. \ref{F:filtration}). They can also be used as an alternative to thresholding a weighted complex, providing a principled approach to binarizing which retains all of the data in the original weighted complex.

In what follows, we avoid introducing technical details beyond those absolutely necessary, as they can be found elsewhere \citep{ghrist2014elementary,nanda2014simplicial,kozlov2007combinatorial}. For readers interested in ways these ideas can be applied to the theory of neural coding, we recommend \citep{curto2016what}.

\section*{Motivating examples}

Before we dive into describing the tools and how they have been used, we begin with a pair of simple thought experiments which highlight more explicitly the reasons we consider these techniques to be valuable for the study of neural systems.

\begin{figure*}
\begin{center}
\includegraphics{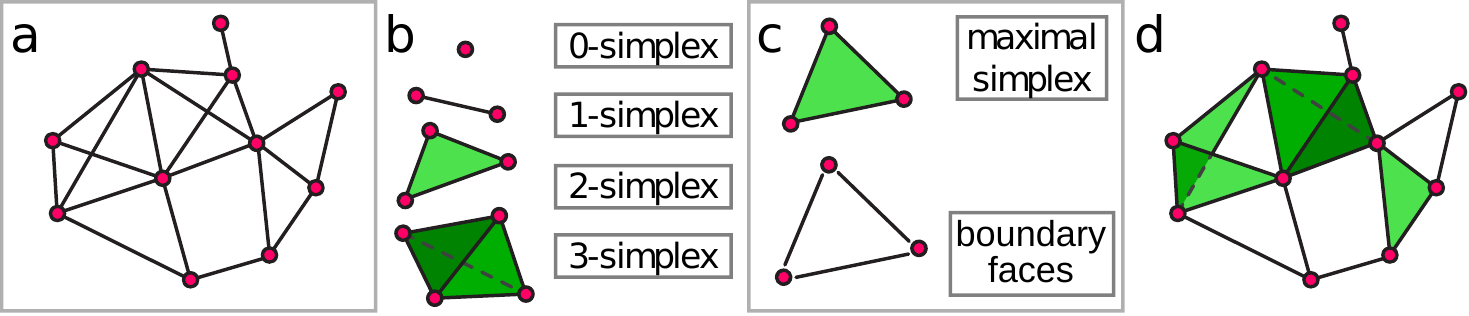}
\end{center}
\caption{Simplicial complexes generalize network models. \emph{(a)} A graph encodes elements of a neural system as vertices and dyadic relations between them as edges. \emph{(b-c)} Simplicial complex terminology. A simplicial complex is made up of \emph{vertices} and \emph{simplices}, which are defined in terms of collections of vertices. \emph{(b)} A $n$-simplex can be thought of as the convex hull of $(n+1)$ vertices. \emph{(c)} The \emph{boundary} of a simplex consists of all possible subsets of its constituent vertices, called its \emph{faces}, which are themselves required to be simplices in the complex. A simplex which is not in the boundary of any other simplex is called \emph{maximal}. \emph{(d)} A simplicial complex encodes polyadic relations through its simplices. Here, in addition to the dyadic relations specified by the edges, the complex specifies one four-vertex relation and three three-vertex relations. The omission of larger simplices where all dyadic relations are present, such as the three bottom-left vertices or the four top-left vertices, encodes structure that cannot be specified using network models. }
\label{F:sim_cx}
\end{figure*}

First, imagine a simple neural system consisting of three brain regions (or neurons) with unknown connectivity. One possible activity profile for such a population includes some sort of sequential information processing loop or ``pacemaker'' like circuit, where the regions activate in a rotating order (Fig. \ref{F:intro}c, top). A second is for all three of the regions to be active simultaneously when engaged in certain computations, and otherwise quiescent or uncorrelated (Fig. \ref{F:intro}c, bottom). In either case, an observer would find the activity of all three possible pairs of regions to be strongly correlated. Because a network can only describe dyadic relationships between population elements, any binary coactivity network constructed from such observations would necessarily be identical for both (Fig. \ref{F:intro}d). However, a more versatile language could distinguish the two by explicitly encoding the triple coactivity pattern in the second example (Fig. \ref{F:intro}e). The framework of \emph{simplicial complexes} (Fig. \ref{F:sim_cx}b-\ref{F:sim_cx}d) is such a language, a straightforward extension of the formalism of graph theory that allows one to describe relations between arbitrarily large sub-populations without sacrificing computability or access to many of the fundamental tools of network science. Further, the richer structure inherent in simplicial complexes has driven the development of correspondingly more powerful mathematical techniques for detecting and analyzing the structure of the systems they encode. These methods provide a quantitative architecture through which to address modern questions about complex and emergent behavior in neural systems.

Second, consider a much larger neural system, consisting of several hundred units, whose activity is summarized as a correlation or coherence matrix (Fig. \ref{F:intro}f, top). It is common practice to binarize such a matrix by thresholding it at some value, taking entries above that value to be ``significant'' connections, and to study the resulting, much sparser network (Fig. \ref{F:intro}f, bottom). Selecting this significance level is problematic, particularly when the underlying system is not thoroughly understood and low-impact effects that might be dismissed as noise are potentially important to its function. One method for working around this difficulty is to take several thresholds and study the results separately. However, this approach still discards most of the information contained in the edge weights, much of which can be of inherent value in understanding the system. We propose instead the use of \emph{filtrations}, which record the results of every possible binarization of the network, along with the associated threshold value (Fig. \ref{F:intro}g). Filtrations not only retain all of the information in the original weighted networks, but unfold that information into a more accessible form, allowing one to lift any measure of structure in networks (or simplicial complexes) to ``second order'' measures as functions of edge weight (Fig. \ref{F:intro}h). Such functions carry information, for example, in their rate of change, where sudden phase transitions in network structure as one varies the threshold can indicate the presence of modules or rich clubs in networks (Fig. \ref{F:intro}h). Alternately, the area under such curves was used in \citep{giusti2015clique} to detect geometric structure in the activity of hippocampal neural populations (Fig. \ref{F:giusti_pnas}). Further, even more delicate information can be extracted from the filtration by tracking the \emph{persistence} of individual structures in the graphs (such as components) as the threshold varies (Fig. \ref{F:filtration}c).

\section*{A Growing Literature}
\begin{figure*}[t]
\begin{center}
\includegraphics[width=0.8\textwidth]{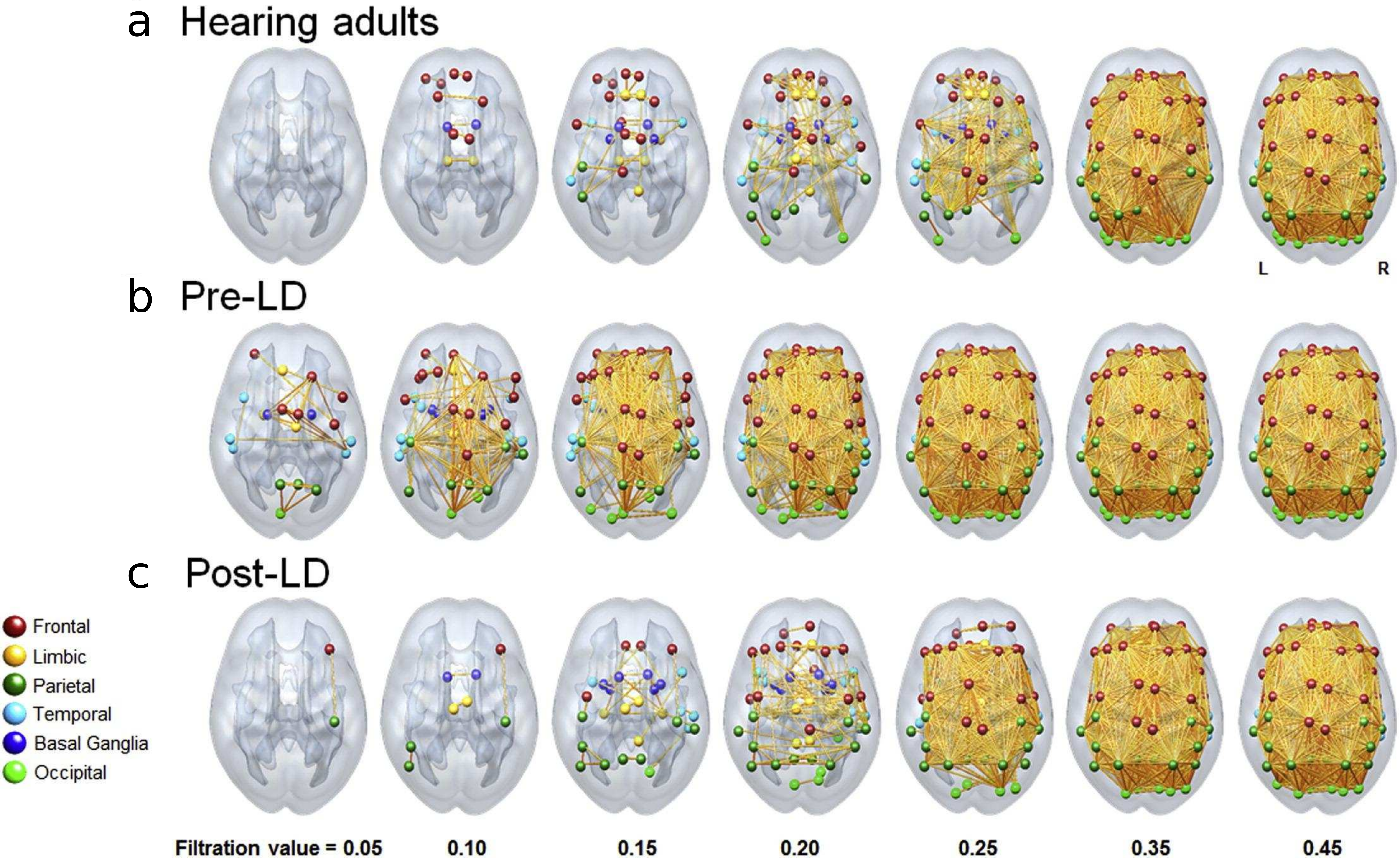}
\end{center}
\caption{Filtered brain networks constructed from interregional correlations of density from MRI detect differences in hearing and deaf populations. Density correlation networks obtained from \emph{(a)} hearing, \emph{(b)} prelingual deaf, and \emph{(c)} postlingual deaf adults. Differences in the evolution of network components across groups as the threshold parameter varies provides insight into differences in structure. It is unclear how one would select a particular threshold which readily reveals these differences without {\it a priori} knowledge of their presence. Figure reproduced with permission from \citep{kim2014morphological}.}
\label{F:hearing_paper}
\end{figure*}

Before we begin a careful discussion of the mathematical concepts described above, we provide an overview of the existing literature, which can be roughly divided into two branches:

\smallskip

\begin{figure*}[t]
\begin{center}
\includegraphics[width=\textwidth]{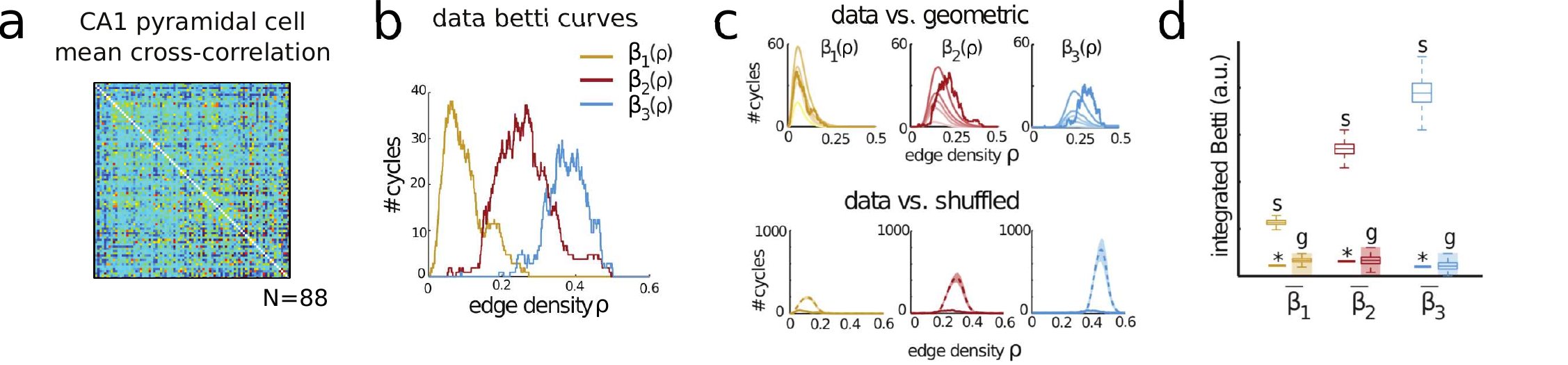}
\end{center}
\caption{Betti numbers detect the existence of geometric organizing principles in neural population activity from rat hippocampus. \emph{(a)} Mean cross correlation of N=88 rat CA1 pyramidal cells activity during spatial navigation. \emph{(b)} Betti numbers as a function of graph edge density (\# edges / possible \# edges) for the clique complex of the pairwise correlation network in (a). \emph{(c)} Comparison of data Betti numbers (thick lines) to model random networks with (top) geometric weights given by decreasing distance between random points in Euclidean space and (bottom) with no intrinsic structure obtained by shuffling the entries of the correlation matrix. \emph{(d)} Integrals of the curves from panel B show that the data (thick bars) lie in the geometric regime (g) and that the unstructured network model (s) is fundamentally different ($p < 0.001$). Similar geometric organization was observed in non-spatial behaviors such as REM sleep. Figure reproduced with permission from \citep{giusti2015clique}.}
\label{F:giusti_pnas}
\end{figure*}

\noindent \textbf{Building simplicial complexes to describe neural coding and network properties.} In \citep{curto2008cell}, a novel kind of simplicial complex derived from neural data was introduced to show how hippocampal place cell activity can, in principle, be used to reconstruct the topology of the represented environment. The fundamental observation is that place fields corresponding to nearby locations will overlap, and thus neurons corresponding to those fields will be co-active (Fig. \ref{F:data}b). Theoretical tools from algebraic topology then imply that (assuming convexity of place fields) one can work backward from a simplicial complex built from these observed coactivity patterns to recover the intersection pattern of the receptive fields, thus describing a \emph{topological map} of the animal's environment. In order to recover the geometry of the environment, one can introduce information regarding receptive field size \citep{curto2008cell}, however it seems plausible that place cells intrinsically record only these intersection patterns and rely on downstream mechanisms for interpretation of such geometry. This hypothesis is supported by an interesting experiment of  \citep{dabaghian2014reconceiving}, in which place cell activity was recorded before and after deformation of segments of the legs of a U-shaped track and shown to be consistent; a geometric map would have been badly deformed by such a change in the environment, while a topological map would remain consistent. Further theoretical and computational work has explored how such topological maps might form \citep{dabaghian2012topological} and shown that theta oscillations improve such learning mechanisms  \citep{arai2014effects}, as well as demonstrating how one might use this understanding to decode maps of the environment from observed cell activity \citep{chen2014neural}.

Even in the absence of an expected underlying collection of spatial receptive fields, similar tools can be employed to explore how network modules interact. In \citep{ellis2014describing}, the authors construct a filtration of simplicial complexes from fMRI recordings, tracking not only which regions were coactive, but how often they were observed to be active together. Such a  filtration provides quantitative tools for detecting computational units, even when those units may change dynamically over time: units will appear very early in the filtration, while coincidental interactions will happen less often and thus appear only much later. The same approach was used in \citep{pirino2014topological}, to differentiate \emph{in vivo} cortical cell cultures into functional sub-networks under various system conditions. Finally, an extension of these ideas which includes a notion of \emph{directedness} has been used to investigate the relationship between simulated structural and functional neural networks \citep{dlotko2016topological}.

\smallskip
\noindent \textbf{Using measurements of filtrations to characterize brain architecture or state.} One of the earliest applications of algebraic topology to neural data was to the study of activity in the macaque primary visual cortex \citep{singh2008topological}, where different distributions of algebraic-topological features provided a mechanism for distinguishing recordings of spontaneous activity from those obtained during exposure to natural images. These features, called \emph{cycles}, provide a measurement of the mesoscale or global structure of the system being studied. 

The presence or absence of cycles can represent many different elements of interest in structural data: in \citep{chung2009persistence}, the authors use the statistics of cycles representing regions of thin cortex to differentiate human ASD subjects from controls; in \citep{brown2012structure}, cycles constructed from physical locations in space are used to understand the spatial structure of afferent neuron terminals in crickets \citep{brown2012structure}; and, in \cite{bendich2014persistent}, the authors use two different types of cycles derived from brain artery trees to detect age and gender in human subjects. 

Also common has been the use of correlation of observed neuronal population activity to construct weighted graphs, from these to construct filtered simplicial complexes and then compute algebraic-topological measurements to be used as discriminators of classes of subjects. Focusing on how components persistent as the filtration parameter varies, this technique was used in \citep{lee2011discriminative} to classify pediatric ADHD, ASD and control subjects; in \citep{khalid2014tracing} to differentiate mouse models of depression from controls; in \citep{choi2014abnormal} to differentiate epileptic rat models from controls; and in  \citep{kim2014morphological} to study morphological correlations in adults with hearing loss (Fig. \ref{F:hearing_paper}). Studying the persistence of more complex cycles computed from fMRI recordings distinguishes subjects under psilocybin condition from controls \citep{petri2014homological}, and a similar approach has been used for the study of functional brain networks during learning \citep{stolz2014computational}. More recently, these techniques have been adapted to detect structure, such as that possessed by a network of hippocampal place cells, in the information encoded by a neural population through observations of its activity without reference to external data such as animal behavior \citep{giusti2015clique} (Fig. \ref{F:giusti_pnas}).

The field of topological neuroscience is both very new and very small, yet it already offers an array of powerful new quantitative approaches for addressing the unique challenges inherent in understanding neural systems and it has begun making substantial contributions. In recent years, there have been a number of innovative collaborations between mathematicians interested in applying topological methods and researchers in a variety of biological disciplines, including the discovery of new genetic markers for breast cancer survival \citep{nicolau2011topology}, measurement of structure and stability of biomolecules \citep{gameiro2013topological,xia2015persistent}, new frameworks for understanding viral evolution \citep{chan2013topology}, and characterization of dynamics in gene regulatory networks \citep{boczko2005structure}. This wide-spread interest is an untapped resource for empirical neuroscientists which promises to facilitate both direct applications of existing techniques and the collaborative construction of novel tools specific to their needs.

\smallskip

We devote the remainder of the paper to a careful exposition of these techniques, highlighting specific ways that they may or have already been used to address questions of interest to neuroscientists.

\section*{Mathematical Framework: Simplicial complexes}

We begin with a short tutorial on simplicial complexes, and illustrate the similarities and differences with graphs.

\setlength{\intextsep}{0pt}%
\begin{figure}[b]
\fbox{\begin{minipage}{\dimexpr\linewidth-2\fboxrule-2\fboxsep}
\textbf{Formal Definitions}\\
An \emph{(abstract) simplicial complex} $X$ is a pair of sets: $V_X,$ called the \emph{vertices}; and $S_X,$ called the \emph{simplices}, each of which is a finite subset of $V_X,$ subject to the requirement that if $\sigma$ is in $S_X$, then every subset $\tau$ of $\sigma$ is also in $S_X$. A simplex with $n$ elements is called an \emph{$(n-1)$-simplex}, and subsets $\tau \subset \sigma$ are \emph{faces} of $\sigma$.
\end{minipage}}
\end{figure}

A \emph{simplicial complex}, like a graph, consists of a set of vertices and a specified collection of subsets of those vertices, called \emph{simplices}, subject to the mild restriction that any subset of a simplex must also be a simplex. Observe that any graph is automatically a simplicial complex with all simplices being either vertices or pairs (edges). General simplicial complexes possess more subtle information.

\begin{table*}
\begin{center}
\begin{tabular}{|p{0.26\linewidth}|p{0.65\linewidth}|}
\hline
Simplicial Complex Type& Utility\\
\hline
Graph & General framework for encoding dyadic relations\\
Clique Complex& Canonical polyadic extension of existing network models\\
Concurrence Complex/Dual& Relationships between two variables of interest\\
& e.g., time and activity, or activity in two separate regions\\
Independence Complex& Structure where non-membership satisfies the simplex property\\
& e.g., communities in a network\\
\hline
\end{tabular}
\end{center}
\caption{Comparison of sample types of simplicial complexes for encoding neural data.}
\end{table*}

Just as one can represent a graph as a collection of points and line segments between them, one can represent the simplices in a simplicial complex as a collection of solid regions connecting vertices (Fig. \ref{F:sim_cx}d). Under this geometric interpretation, a single vertex is a zero-dimensional point, while two distinct points define a one-dimensional line segment, three points a two-dimensional triangle, and so on. Terminology for simplices is derived from this geometric representation: a simplex on $(n+1)$ vertices is called an $n$-simplex and is viewed as spanning an $n$-dimensional region. Further, the requisite subsets of a simplex represent regions in the geometric boundary of the simplex (Fig. \ref{F:sim_cx}e), so these subsets of a simplex are called its \emph{faces}.

Because any given simplex is required to ``contain all of its faces'', to identify a complex it is sufficient to specify only the \emph{maximal simplices}, those which do not appear as faces of another simplex (Fig. \ref{F:sim_cx}e). This  dramatically reduces the amount of data necessary for working with simplicial complexes, which helps make computations feasible.

In real-world systems, simplicial complexes possess richly structured patterns that can be detected and characterized using recently developed computational tools from algebraic topology \citep{carlsson2009topology,lum2013extracting}, just as graph theoretic tools can be used to study networks. Importantly, these tools reveal much deeper properties of the relationships between vertices than graphs, and many are constructed not only to see structure in individual simplicial complexes, but also to help one understand how two or more simplicial complexes compare or relate to one another. These capabilities naturally enable the study of complex dynamic structure in neural systems, and formalize statistical inference via comparisons to null models.

\section*{How do we encode neural data?}

To demonstrate the broad utility of this framework, we turn to describing a selection of the many types of simplicial complexes that can be constructed from data: the clique complex
, the concurrence complex \citep{ellis2014describing,curto2008cell,dowker1952homology}, its Dowker dual \citep{dowker1952homology}, and the independence complex \citep{kozlov2007combinatorial}. In each case, we describe the relative utility in representing different types of neural data -- from spike trains measured from individual neurons to BOLD activations measured from large-scale brain areas.

\begin{figure*}
\begin{center}
\includegraphics{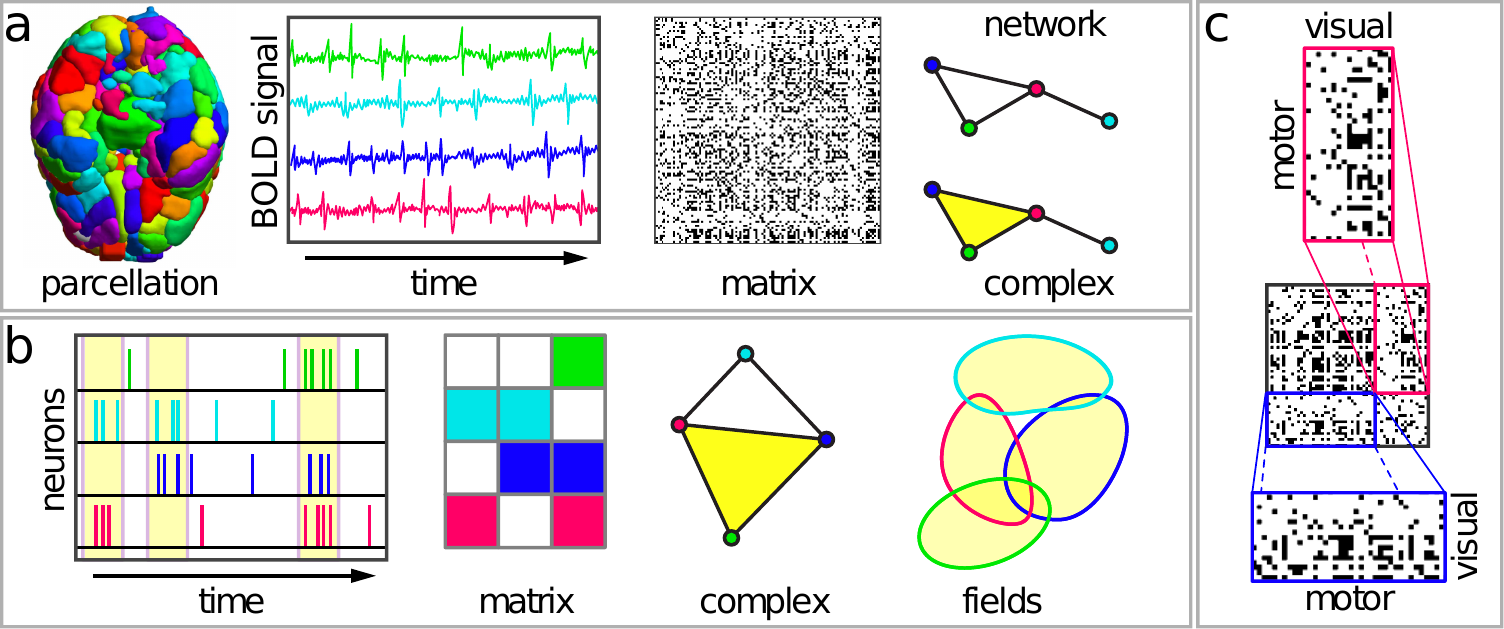}
\end{center}

\caption{Simplicial complexes encode diverse neural data modalities. \emph{(a)} Correlation or coherence matrices between regional BOLD time series can be encoded as a type of simplicial complex called a \emph{clique complex}, formed by taking every complete (all-to-all) subgraph in a binarized functional connectivity matrix to be a simplex. \emph{(b)} Coactivity patterns in neural recordings can be encoded as a type of simplicial complex called a \emph{concurrence complex}. Here, we study a binary matrix in which each row corresponds to a neuron and each column corresponds to a collection of neurons that is observed to be coactive at the same time (yellow boxes) -- i.e., a simplex. \emph{(c)} Thresholded coherence between the activity patterns of motor regions and visual regions in human fMRI data during performance of a motor-visual task \citep{Bassett2013a}. (top) We can construct a concurrence complex whose vertices are motor regions and whose simplices are families of motor regions whose activity is strongly coherent with a given visual region. (bottom) We can also construct a \emph{dual} complex whose vertices are families of motor regions. The relationship between these two complexes carries a great deal of information about the system \citep{dowker1952homology}.}
\label{F:data}
\end{figure*}

\noindent \textbf{Clique Complex.} One straightforward method for constructing simplicial complexes begins with a graph where vertices represent neural units and edges represent structural or functional connectivity between those units (Fig. \ref{F:sim_cx}a-\ref{F:sim_cx}b). Next, one replaces every {\em clique} (all-to-all connected subgraph) by a simplex on the vertices participating in the clique (Fig. \ref{F:data}a). This procedure produces a \emph{clique complex}, which encodes the same information as the underlying graph, but additionally completes the skeletal network to the fullest simplicial structure. The utility of this structure was recently demonstrated in the context of neural activity measured in rat hippocampal pyramidal cells during both spatial and non-spatial behavior (including REM sleep) \citep{giusti2015clique} (Fig. \ref{F:giusti_pnas}). In contrast to graph statistics, the pattern of simplices revealed the presence of geometric structure in only the information encoded in neural population activity correlations that -- surprisingly -- could be identified and characterized independently from the animal's position. This application demonstrates that simplicial complexes are sensitive to organizational principles that are hidden to graph statistics, and can be used to infer parsimonious rules for information encoding in neural systems.

\bigskip
Clique complexes precisely encode the topological features present in a graph. However, other types of simplicial complexes can be used to represent information that \emph{cannot} be so encoded in a graph. 
\bigskip

\noindent \textbf{Concurrence Complex.} Using cofiring, coactivity, or connectivity as before, let us consider relationships between two different sets of variables. For example, we can consider (i) neurons and (ii) times, where the relationship is given by a neuron firing in a given time (Fig. \ref{F:data}b) \citep{curto2008cell}; a similar framing exists for (i) brain regions and (ii) times, where the relationship is given by a brain region being active at a given time \citep{ellis2014describing}. Alternatively, we can consider (i) brain regions in the motor system and (ii) brain regions in the visual system, where the relationship is given by a motor region displaying similar BOLD activity to a visual region (Fig. \ref{F:data}c) \citep{Bassett2015}. In each case, we can record the patterns of relationships between the two sets of variables as a binary matrix, where the rows represent elements in one of the variables (e.g., neurons) and the columns the other (e.g., times), with non-zero entries corresponding to the row-elements in each column sharing a relation (e.g., firing together at a single time). The \emph{concurrence complex} is formed by taking the rows of such a matrix as vertices and the columns to represent maximal simplices consisting of those vertices with non-zero entries \citep{dowker1952homology}. A particularly interesting feature of this complex is that it remains naive to coactivity patterns that do not appear, and this naivety plays an important role in its representational ability; for example, such a complex can be used to decode the geometry of an animal's environment from observed hippocampal cell activity \citep{curto2008cell}.

\bigskip

Moving to simplicial complex models provides a dramatically more flexible framework for specifying data encoding than simply generalizing graph techniques. Here we describe two related simplicial complex constructions from neural data which cannot be represented using network models.

\bigskip
\noindent \textbf{Dowker Dual.}
Beginning with observations of coactivity, connection or cofiring as before, one can choose to represent neural units as simplices whose constituent vertices represent patterns of coactivity in which the unit participates. Expressing such a structure as a network would necessitate every neural unit participating in precisely two activity patterns, an unrealistic requirement, but is straightforward in the simplicial complex formalism. Mathematically speaking, one can think of the matrix encoding this complex as the transpose of the matrix encoding the concurrence complex; such ``dual'' complexes are deeply related to one another, as first observed in \citep{dowker1952homology}. Critically, this formulation refocuses attention (and the output of various vertex-based statistical measures) from individual neural units to patterns of coactivity.
\bigskip

\noindent \textbf{Independence Complex.} It is sometimes the case that an observed structure does not satisfy the simplicial complex property, but its complement does. One example of interest is the collection of \emph{communities} in a network \citep{Fortunato2010,Porter2009}: communities are subgraphs of a network whose vertices are more densely connected to one another than expected in an appropriate null model. The collection of vertices in the community is not a simplex, because removing densely connected vertices can cause the community to dissolve. Thus, community structure is well-represented as a \emph{hypergraph} \citep{Bassett2014}, though such structures are much less natural and useful than simplicial complexes. However, in this setting simplices can be taken to be all vertices \emph{not} in a community. Such a simplicial complex is again essentially a concurrence complex: simply negate the binary matrix whose rows are elements of the network and columns correspond to community membership. Such a complex is called an \emph{independence complex} \citep{kozlov2007combinatorial}, and can be used to study properties of a system's community structure such as dynamic flexibility \citep{Bassett2011b,Bassett2013a}.

Together, these different types of complexes can be used to encode a wide variety of relationships (or lack thereof) among neural units or coactivity properties in a simple matrix that can be subsequently interrogated mathematically. This is by no means an exhaustive list of complexes of potential interest to the neuroscience community; for further examples, we recommend \citep{ghrist2014elementary,kozlov2007combinatorial}.

\section*{How do we measure the structure of simplicial complexes?}

\begin{figure*}
\begin{center}
\includegraphics{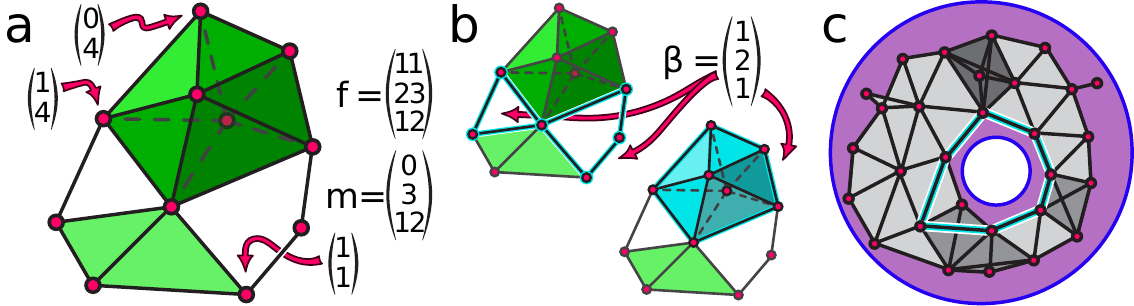}
\end{center}
\caption{Quantifying the structure of a simplicial complex. \emph{(a)} Generalizations of the degree sequence for a simplicial complex. Each vertex has a \emph{degree} vector giving the number of maximal simplices of each degree to which it is incident. The \emph{f-vector} gives a list of how many simplices of each degree are in the complex, and the \emph{maximal simplex distribution} records only the number of maximal simplices of each dimension. \emph{(b)} Closed cycles of dimension 1 and 2 in the complex from panel A. (left) There are two independent 1-cycles (cyan) up to deformation through 2-simplices, and (right) a single 2-cycle (cyan) enclosing a 3-d volume. The Betti number vector $\beta$ gives an enumeration of the number of $n$-cycles in the complex, here with $n=0, 1$ and $2$; the single 0-cycle corresponds to the single connected component of the complex. \emph{(c)} Schematic representation of the reconstruction of the presence of an obstacle in an environment using a concurrence complex constructed from place cell cofiring \citep{curto2008cell}. By choosing an appropriate cofiring threshold, based on approximate radii of place cell receptive fields, there is a single 1-cycle (cyan), up to deformation through higher simplices, indicating a large gap in the receptive field coverage where the obstacle appears.}
\label{F:structure}
\end{figure*}

Just as with network models, once we have effectively encoded neural data in a simplicial complex, it is necessary to find useful quantitative measurements of the resulting structure to draw conclusions about the neural system of interest. Because simplicial complexes generalize graphs, many familiar graph statistics can be extended in interesting ways to simplicial complexes. However, algebraic topology also offers a host of novel and very powerful tools that are native to the class of simplicial complexes, and cannot be derived from well known graph theoretical constructs.

\noindent \textbf{Graph Theoretical Extensions.} First, let us consider how we can generalize familiar graph statistics to the world of simplicial complexes. The simplest local measure of structure -- the \emph{degree} of a vertex -- naturally becomes a vector-measurement whose entries are the number of maximal simplices of each size in which the vertex participates (Dlotko et al., unpublished) (Fig. \ref{F:structure}a). Although a direct extension of the degree, this vector is perhaps more intuitively thought of as a generalization of the \emph{clustering coefficient} of the vertex: in this setting we can distinguish empty triangles, which represent three dyadic relations but no triple-relations, from 2-simplices which represent clusters of three vertices (and similarly for larger simplices).

Just as we can generalize the degree, we can also generalize the degree distribution. Here, the \emph{simplex distribution} or \emph{f-vector} is the global count of simplices by size, which provides a global picture of how tightly connected the vertices are; the \emph{maximal simplex distribution} collects the same data for maximal faces (Fig. \ref{F:structure}a). While these two measurements are related, their difference occurs in the complex patterns of overlap between simplices and so together they contain a great deal of structural information about the simplicial complex. Other local and global statistics such as \emph{efficiency} and \emph{path length} can be generalized by considering paths through simplices of some fixed size, which provides a notion of \emph{robust connectivity} between vertices of the system \citep{dlotko2016topological}; alternately, a path through general simplices can be assigned a strength coefficient depending on the size of the maximal simplices through which it passes.

\noindent \textbf{Algebraic-Topological Methods.} Such generalizations of graph-theoretic measures are possible, and likely of significant interest to the neuroscience community, however they are not the fundamental statistics originally developed to characterize simplicial complexes. In their original context, simplicial complexes were used to study shapes, using \emph{algebraic topology} to measure global structure. Thus, this framework also provides new and powerful ways to measure biological systems.

The most commonly used of these measurements is the \emph{(simplicial) homology} of the complex\footnote{Names of topological objects have a seemingly pathological tendency to conflict with terms in biology, so long have the two subjects been separated. Mathematical homology has no \emph{a priori} relationship to the usual biological notion of homology.}, which is actually a sequence of measurements. The \emph{$n^{\rm th}$ homology} of a simplical complex is the collection of \emph{(closed) $n$-cycles} formed from $n$-simplices (Fig. \ref{F:structure}b), up to a notion of equivalence\footnote{Two $n$-cycles are equivalent if they differ by the boundary of some collection of $(n+1)$-simplices.}. Such cycles can be thought of as characterizing ``holes'' in various dimensions\footnote{The actual definition of a cycle is more subtle and requires careful discussion. We refer the interested reader to the aforementioned expositions \citep{ghrist2014elementary,nanda2014simplicial}.}, and are an example of global structure arising from local structure; simplices arrayed across multiple vertices must coalesce in a particular fashion to encircle a hole not filled in by other simplices. In many settings, a powerful summary statistic is simply a count of the number of inequivalent cycles of each dimension appearing in the complex.  These counts are called \emph{Betti numbers}, and collect them as a vector $\beta$.

\setlength{\intextsep}{0pt}%
\begin{figure}
\fbox{\begin{minipage}{\dimexpr\linewidth-2\fboxrule-2\fboxsep}
\textbf{Formal Definitions}\\
In order to compute with simplicial complexes, we convert their assembly implicit assembly instructions into linear algebra as follows. For a simplicial complex $X$, define a sequence of vector spaces $C_n(X)$ (over the finite field $\mathbb{F}_2=\{0,1\}$) with bases the $n$-simplices of $X$, writing $[\sigma]$ for the basis element corresponding to the simplex $\sigma$. Then define the \emph{boundary} maps as linear transformations $\partial_n\colon C_n(X) \rightarrow C_{n-1}(X)$ assigning to each $n$-simplex $[\sigma]$ the formal sum $\sum_[\tau]$ over all $(n-1)$-simplex faces $[\tau]$ of $\sigma$. The \emph{$n$th homology} of $X$ is defined as the quotient vector space $H_n(X) = \text{ker}\;\partial_n \,/\, \text{im}\;\partial_{n+1}$, and the \emph{$n$th Betti number} is its dimension $\beta_n = \text{dim}\,H_n(X)$.
\end{minipage}}
\end{figure}

In the context of neural data, the presence of multiple homology classes indicates potentially interesting structure whose interpretation depends on the meaning of the vertices and simplices in the complex. For example, the open triangle in the complex of Fig. \ref{F:data}b is a 1-cycle representing pairwise coactivity of all of the constituent neurons but a lack of triple coactivity; thus, the reconstructed receptive field model includes no corresponding triple intersection, indicating a hole or obstacle in the environment. In the context of regional coactivity in fMRI, such a 1-cycle might correspond to observation of a distributed computation that does not involve a central hub. Cycles of higher dimension are more intricate constructions, and their presence or absence can be used to detect a variety of other more complex, higher-order features.

\section*{Additional Tools to Assess Hierarchical and Temporal Structure}

In previous sections we have seen how we can construct simplicial complexes from neural data and interrogate the structure in these complexes using both extensions of common graph theoretical notions and completely novel tools drawn from algebraic topology. We close the mathematical portion of this exposition by discussing a computational process that is common in algebraic topology and that directly addresses two critical needs in the neuroscience community: (i) the assessment of hierarchical structure in relational data via a principled thresholding approach, and (ii) the assessment of temporal properties of stimulation, neurodegenerative disease, and information transmission.

\setlength{\intextsep}{0pt}%
\begin{figure}
\fbox{\begin{minipage}{\dimexpr\linewidth-2\fboxrule-2\fboxsep}
\textbf{Formal Definitions}\\
A \emph{filtration} of a simplicial complex $X$ is a sequence of simplicial subcomplexes of the form
\begin{equation*}
\emptyset = X_0 \subset X_1 \subset X_2 \subset \cdots \subset  X_M=X,
\end{equation*}
where $\subset$ denotes inclusion as a subcomplex. 

A simplicial complex is \emph{weighted} if each simplex $\sigma$ is assigned a real-valued weight $w(\sigma)$ so that if $\tau$ is a face of $\sigma$, then $w(\tau) \geq w(\sigma)$. Every weighted simplicial complex can be naturally converted into a filtration. Because the complex contains a finite number of simplices, the weight function $w$ takes on only finitely many values, $w_1 \geq w_2 \geq \cdots w_M$. We can thus construct a subcomplex $X_i$ of $X$ by considering the subcomplexes of $X$ for which $w(\sigma) \geq w_i$. These subcomplexes fit together to form a filtration (Fig. \ref{F:filtration}B).
\end{minipage}}
\end{figure}

\noindent \textbf{Filtrations to Assess Hierarchical Structure in Weighted Networks.} One of the most common features of network data is a notion of \emph{strength} or \emph{weight} of connections between nodes. In some situations, like measurements of correlation or coherence of activity, the resulting network has edges between every pair of nodes and it is common to \emph{threshold} the network to obtain some sparser, unweighted network whose edges correspond to ``significant'' connections \citep{Achard2006}. However it is difficult to make a principled choice of threshold \citep{Ginestet2011,Bassett2012a,Garrison2015,Drakesmith2015,Sala2014,Langer2013}, and the resulting network discards a great deal of information. Even in the case of sparse weighted networks, many metrics of structure are defined only for the underlying unweighted network, so in order to apply the metric, the weights are discarded and this information is again lost \citep{Rubinov2011}.

\begin{figure*}
\begin{center}
\includegraphics{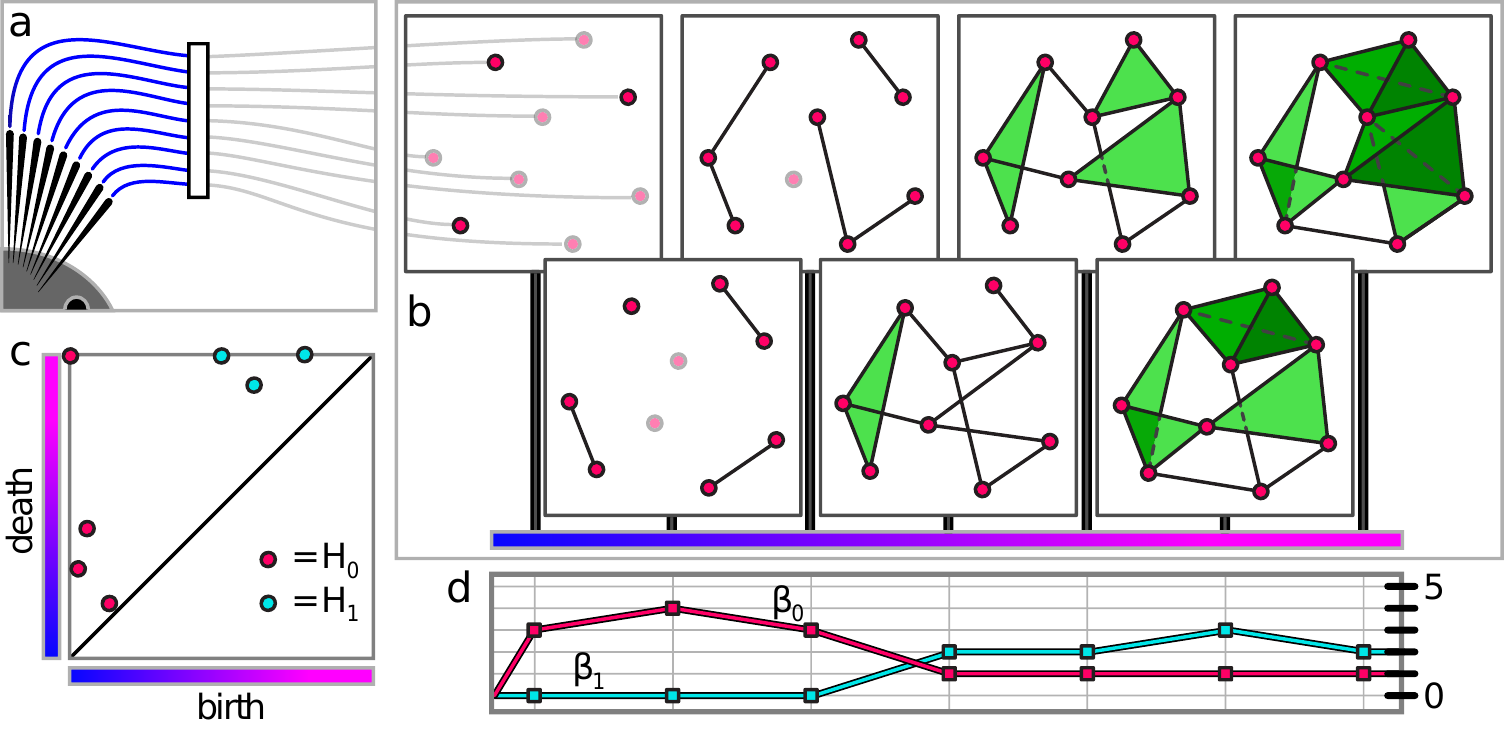}
\end{center}
\caption{Filtrations of a weighted simplicial complex measure dynamic network properties. \emph{(a)} A neural system can be stimulated in precise locations using electrical, magnetic or optogenetic methods and the resulting activity recorded. \emph{(b)} A filtration of simplicial complexes is built by recording as maximal faces all patterns of coactivity observed up to a given time. Filtrations can be constructed from any weighted simplicial complex by thresholding using some \emph{filtration parameter}. \emph{(c)} A \emph{persistence diagram} recording the appearance (``birth'') and disappearance or merging (``death'') of homology cycles throughout the filtration in panel (b).  Cycles on the top edge of the diagram are those that do not die. Tracking equivalent cycles through the filtration provides information about the evolution of structure as the filtration parameter changes. \emph{(d)} \emph{Betti curves} are the Betti numbers for each complex in the filtration of panel (b) represented as functions of time. Such curves can be constructed for any numerical measurement of the individual unweighted simplicial complexes in the filtration and provide a more complete description of structure than the individual measurements taken separately.}
\label{F:filtration}
\end{figure*}

Here, we describe a technique that is commonly applied in the study of weighted simplicial complexes which does not discard any information. Generalizing weighted graphs, {\it weighted simplicial complexes} have assigned to each of simplex (including vertices) a numeric \emph{weight}, subject to the restriction that the weight of a simplex is no larger than that of any of its faces. That is, an observed relationship between any subset of a population is at least as strong as that observed among any larger subpopulation containing it. Given a weighted simplicial complex, a \emph{filtration} of complexes can be constructed by consecutively applying each of the weights as thresholds and labeling each complex by the weight at which it was binarized. The resulting sequence of complexes retains all of the information in the original weighted complex, but one can apply metrics that are undefined or difficult to compute for weighted complexes to the entire collection, thinking of the resulting values as a function parameterized by the weights of the original complex (Fig. \ref{F:filtration}d). However, it is also the case that these unweighted complexes are related to one another, and more sophisticated measurements of structure, like homology, can exploit these relations to extract much finer detail of the evolution of the complexes as the threshold varies (Fig. \ref{F:filtration}c). We note that the omni-thresholding approach utilized in constructing a filtration is a common theme among other recently developed methods for network characterization, including cost integration \citep{Ginestet2011} and functional data analysis \citep{Bassett2012a,ellis2014describing}.

\bigskip
The formalism described above provides a principled framework to translate a weighted graph or simplicial complex into a family of unweighted graphs or complexes that retain all information in the weighting by virtue of their relationships to one another. However, filtrations are much more generally useful: for example, they can be used to assess the dynamics of neural processes.

\noindent \textbf{Filtrations to Assess Temporal Dynamics of Neural Processes in Health and Disease}

Many of the challenges faced by cutting edge experimental techniques in the field of neuroscience are driven by the underlying difficulties implicit in assessing temporal changes in complex patterns of relationships. For example, with new optogenetics capabilities, we can stimulate single neurons or specific groups of neurons to control their function \citep{Grosenick2015}. Similarly, advanced neurotechnologies including microstimulation, transcranial magnetic stimulation, and neurofeedback enable effective control over larger swaths of cortex \citep{Krug2015,Sulzer2013}. With the advent of these technologies, it becomes imperative to develop computational tools to quantitatively characterize and assess the impact of stimulation on system function, and more broadly, to understand how the structure of a simplicial complex affects the transmission of information.

To meet this need, one can construct a different type of filtration, such as that introduced in \citep{taylor2015topological} in the context of graphs: construct a sequence of simplicial complexes with a time parameter, labeling each simplex as ``on'' or ``off'' at each time,  and require that once simplices ``turn on'' they remain so indefinitely. If the function has the further requirement that in order for a simplex to be active, all of its faces must be as well, then a filtration is obtained by taking all active simplices at each time. Such functions are quite natural to apply to the study of the pattern of neurons or neural units that are activated following stimulation.

Interestingly, this type of filtration is also a natural way in which to probe and reason about models of neurodegenerative disease such as the recently posited \emph{diffusion model} of fronto-temporal dementia \citep{Raj2012,Zhou2012}. Here, critical network epicenters form points of vulnerability that are effected early in the disease, and from which toxic protein species travel via a process of transneuronal spread.  Indeed, these filtrations were first introduced in the context of contagion models \citep{taylor2015topological}, where a simplex becomes active once sufficiently many nearby simplices are active.

\noindent \textbf{Measuring the Structure of Filtrations}

Assuming we have encoded our data in an appropriate filtration, guided by our scientific hypothesis of interest, we might next wish to quantitatively characterize and measure the structure in those filtrations. It is important to note that any given measure of the structure of a simplicial complex can be applied to each complex in a filtration in turn, producing a function from the set of weights appearing in the complex to the set of values the measure can take (Fig. \ref{F:filtration}d). This function is a new measure of the structure of the complex which does not rely on thresholds and can highlight interesting details that would not be apparent at any fixed threshold (or small range of thresholds), as well as being more robust to perturbations in the weights than measurements of any individual complex in the filtration.

Of particular interest in this setting are those quantitative measures whose evolution can be explicitly understood in terms of the underlying map of complexes, as then we can exploit the sequence of maps in the filtration to gain a more refined picture of the structure present in the weighted complex. Central among these in terms of current breadth of application and computability is \emph{persistent homology}, which extends the homology of individual complexes to filtrations by tracking how equivalent cycles evolve through the growing filtration. Increasing the parameter sends each cycle to some cycles in the next simplicial complex in the filtration. Therefore, the sequence of subcomplexes in the filtration is transformed by homology into an inter-related family of evolving cycles. Inside this sequence, cycles have well-defined \emph{birth} and \emph{death} weights, between which they evolve as new simplices are added, changing their form. This information is often encoded in \emph{persistence diagrams} for each degree $n$ (Fig. \ref{F:filtration}c), which give a schematic overview of where the cycles are born and die. Understanding these \emph{persistence lifetimes} of cycles can provide critical information about how the system is arranged. These techniques have been applied to uncover structure in the space of natural images \citep{carlsson2008local}, to detect subject gender from the shape of brain artery trees \citep{bendich2014persistent}, and to identify statistically anomolous multi-region activation patterns in fMRI recordings \citep{ellis2014describing}.

\section*{Conclusion}

We sit at a unique juncture in time, in which it is critical to support the principled development of novel computational tools that are not merely modular, but instead are tuned to address specific neuroscientific challenges at hand. With the feverish rise of data being collected from neural systems across species and spatial scales, mathematicians and experimental scientists must necessarily engage in deeper conversation about how meaning can be drawn from minutia. Such conversations will inevitably turn to the common understanding in the neurosciences that it is not necessarily the individual objects of study themselves, but their relations to one another, that provide the real structure of human and animal thought. Though originally developed for entirely different purposes, the algebraic topology of simplicial complexes provides a quantitative methodology uniquely suited to address these needs.

\bibliographystyle{spbasic}

\begin{thebibliography}{62}
\providecommand{\natexlab}[1]{#1}
\providecommand{\url}[1]{{#1}}
\providecommand{\urlprefix}{URL }
\expandafter\ifx\csname urlstyle\endcsname\relax
  \providecommand{\doi}[1]{DOI~\discretionary{}{}{}#1}\else
  \providecommand{\doi}{DOI~\discretionary{}{}{}\begingroup
  \urlstyle{rm}\Url}\fi
\providecommand{\eprint}[2][]{\url{#2}}

\bibitem[{Achard et~al(2006)Achard, Salvador, Whitcher, Suckling, and
  Bullmore}]{Achard2006}
Achard S, Salvador R, Whitcher B, Suckling J, Bullmore E (2006) A resilient,
  low-frequency, small-world human brain functional network with highly
  connected association cortical hubs. {\em Journal of Neuroscience, 26}(1):63--72

\bibitem[{Arai et~al(2014)Arai, Brandt, and Dabaghian}]{arai2014effects}
Arai M, Brandt V, Dabaghian Y (2014) The effects of theta precession on spatial
  learning and simplicial complex dynamics in a topological model of the
  hippocampal spatial map. {\em PLoS Computational Biology, 10}(6)

\bibitem[{Bassett and Bullmore(2006)}]{Bassett2006b}
Bassett DS, Bullmore ET (2006) Small-world brain networks. {\em Neuroscientist,
  12}:512--523

\bibitem[{Bassett et~al(2011)Bassett, Wymbs, Porter, Mucha, Carlson, and
  Grafton}]{Bassett2011b}
Bassett DS, Wymbs NF, Porter MA, Mucha PJ, Carlson JM, Grafton ST (2011)
  Dynamic reconfiguration of human brain networks during learning. {\em Proceedings of the National
  Academy of the Sciences of the United States of America, 108}(18):7641--7646

\bibitem[{Bassett et~al(2012)Bassett, Nelson, Mueller, Camchong, and
  Lim}]{Bassett2012a}
Bassett DS, Nelson BG, Mueller BA, Camchong J, Lim KO (2012) Altered resting
  state complexity in schizophrenia. {\em Neuroimage, 59}(3):2196--207

\bibitem[{Bassett et~al(2013)Bassett, Wymbs, Rombach, Porter, Mucha, and
  Grafton}]{Bassett2013a}
Bassett DS, Wymbs NF, Rombach MP, Porter MA, Mucha PJ, Grafton ST (2013)
  Task-based core-periphery structure of human brain dynamics. {\em PLoS Computational Biology,
  9}(9):e1003,171

\bibitem[{Bassett et~al(2014)Bassett, Wymbs, Porter, Mucha, and
  Grafton}]{Bassett2014}
Bassett DS, Wymbs NF, Porter MA, Mucha PJ, Grafton ST (2014) Cross-linked
  structure of network evolution. {\em Chaos, 24}:013,112

\bibitem[{Bassett et~al(2015)Bassett, Yang, Wymbs, and Grafton}]{Bassett2015}
Bassett DS, Yang M, Wymbs NF, Grafton ST (2015) Learning-induced autonomy of
  sensorimotor systems. {\em Nature Neuroscience, 18}(5):744--751

\bibitem[{Bendich et~al(2014)Bendich, Marron, Miller, Pieloch, and
  Skwerer}]{bendich2014persistent}
Bendich P, Marron J, Miller E, Pieloch A, Skwerer S (2014) Persistent homology
  analysis of brain artery trees. {\em Annals of Applied Statistics} to appear

\bibitem[{Boczko et~al(2005)Boczko, Cooper, Gedeon, Mischaikow, Murdock,
  Pratap, and Wells}]{boczko2005structure}
Boczko EM, Cooper TG, Gedeon T, Mischaikow K, Murdock DG, Pratap S, Wells KS
  (2005) Structure theorems and the dynamics of nitrogen catabolite repression
  in yeast. {\em Proceedings of the National Academy of Sciences of the United
  States of America, 102}(16):5647--5652

\bibitem[{Brown and Gedeon(2012)}]{brown2012structure}
Brown J, Gedeon T (2012) Structure of the afferent terminals in terminal
  ganglion of a cricket and persistent homology. {\em PLoS ONE, 7}(5)

\bibitem[{Bullmore and Sporns(2009)}]{Bullmore2009}
Bullmore E, Sporns O (2009) Complex brain networks: {G}raph theoretical
  analysis of structural and functional systems. {\em Nature Reviews Neuroscience,
  10}(3):186--198

\bibitem[{Bullmore and Bassett(2011)}]{Bullmore2011}
Bullmore ET, Bassett DS (2011) Brain graphs: graphical models of the human
  brain connectome. {\em Annual Reviews Clinical Psychology, 7}:113--140

\bibitem[{Carlsson(2009)}]{carlsson2009topology}
Carlsson G (2009) Topology and data. {\em Bulletin of the American Mathematical
  Society, 46}(2):255--308

\bibitem[{Carlsson et~al(2008)Carlsson, Ishkhanov, De~Silva, and
  Zomorodian}]{carlsson2008local}
Carlsson G, Ishkhanov T, De~Silva V, Zomorodian A (2008) On the local behavior
  of spaces of natural images. {\em International Journal of Computer Vision,
  76}(1):1--12

\bibitem[{Chan et~al(2013)Chan, Carlsson, and Rabadan}]{chan2013topology}
Chan JM, Carlsson G, Rabadan R (2013) Topology of viral evolution. {\em Proceedings
  of the National Academy of Sciences of the United States of America, 110}(46):18,566--18,571

\bibitem[{Chen et~al(2014)Chen, Gomperts, Yamamoto, and
  Wilson}]{chen2014neural}
Chen Z, Gomperts SN, Yamamoto J, Wilson MA (2014) Neural representation of
  spatial topology in the rodent hippocampus. {\em Neural Computation, 26}(1):1--39

\bibitem[{Choi et~al(2014)Choi, Kim, Kang, Lee, Im, Kim, Chung, Lee
  et~al}]{choi2014abnormal}
Choi H, Kim YK, Kang H, Lee H, Im HJ, Kim EE, Chung JK, Lee DS, et~al (2014)
  Abnormal metabolic connectivity in the pilocarpine-induced epilepsy rat
  model: a multiscale network analysis based on persistent homology. {\em NeuroImage,
  99}:226--236

\bibitem[{Chung et~al(2009)Chung, Bubenik, and Kim}]{chung2009persistence}
Chung MK, Bubenik P, Kim PT (2009) Persistence diagrams of cortical surface
  data. {\em In: Information Processing in Medical Imaging}, Springer, pp 386--397

\bibitem[{Crossley et~al(2013)Crossley, Mechelli, V\'{e}rtes, Winton-Brown,
  Patel, Ginestet, McGuire, and Bullmore}]{Crossley2013}
Crossley NA, Mechelli A, V\'{e}rtes PE, Winton-Brown TT, Patel AX, Ginestet CE,
  McGuire P, Bullmore ET (2013) Cognitive relevance of the community structure
  of the human brain functional coactivation network.{\em  Proceedings of the National Academy of the Sciences of the United States of America,
  110}(28):11,583--11,588

\bibitem[{Curto(2016)}]{curto2016what}
Curto C (2016) What can topology tell us about the neural code? forthcoming

\bibitem[{Curto and Itskov(2008)}]{curto2008cell}
Curto C, Itskov V (2008) Cell groups reveal structure of stimulus space. {\em PLoS
  Computational Biology, 4}(10):e1000,205

\bibitem[{Dabaghian et~al(2012)Dabaghian, M{\'e}moli, Frank, and
  Carlsson}]{dabaghian2012topological}
Dabaghian Y, M{\'e}moli F, Frank L, Carlsson G (2012) A topological paradigm
  for hippocampal spatial map formation using persistent homology. {\em PLoS Computational
  Biology, 8}(8):e1002,581

\bibitem[{Dabaghian et~al(2014)Dabaghian, Brandt, and
  Frank}]{dabaghian2014reconceiving}
Dabaghian Y, Brandt VL, Frank LM (2014) Reconceiving the hippocampal map as a
  topological template. {\em Elife, 3}:e03,476

\bibitem[{Dlotko et~al(2016)Dlotko, Hess, Levi, Nolte, Reimann, Scolamiero,
  Turner, Muller, and Markram}]{dlotko2016topological}
Dlotko P, Hess K, Levi R, Nolte M, Reimann M, Scolamiero M, Turner K, Muller E,
  Markram H (2016) Topological analysis of the connectome of digital
  reconstructions of neural microcircuits. arXiv:160101580 [q-bioNC]

\bibitem[{Dowker(1952)}]{dowker1952homology}
Dowker CH (1952) Homology groups of relations. {\em, Annals of Mathematics} pp 84--95

\bibitem[{Drakesmith et~al(2015)Drakesmith, Caeyenberghs, Dutt, Lewis, David,
  and Jones}]{Drakesmith2015}
Drakesmith M, Caeyenberghs K, Dutt A, Lewis G, David AS, Jones DK (2015)
  Overcoming the effects of false positives and threshold bias in graph
  theoretical analyses of neuroimaging data. {\em NeuroImage, 118}:313--333

\bibitem[{Ellis and Klein(2014)}]{ellis2014describing}
Ellis SP, Klein A (2014) Describing high-order statistical dependence using
  “concurrence topology,” with application to functional mri brain data. {\em Homology, Homotopy and Applications, 16}(1)

\bibitem[{Feldt et~al(2011)Feldt, Bonifazi, and Cossart}]{Feldt2011}
Feldt S, Bonifazi P, Cossart R (2011) Dissecting functional connectivity of
  cortical microcircuits: experimental and theoretical insights. {\em Trends in
  Neurosciences, 34}:225--236

\bibitem[{Fortunato(2010)}]{Fortunato2010}
Fortunato S (2010) Community detection in graphs. {\em Physics Reports, 486}(3--5):75--174

\bibitem[{Gameiro et~al(2013)Gameiro, Hiraoka, Izumi, Kramar, Mischaikow, and
  Nanda}]{gameiro2013topological}
Gameiro M, Hiraoka Y, Izumi S, Kramar M, Mischaikow K, Nanda V (2013) A
  topological measurement of protein compressibility. {\em Japan Journal of
  Industrial and Applied Mathematics, 32}(1):1--17

\bibitem[{Garrison et~al(2015)Garrison, Scheinost, Finn, Shen, and
  Constable}]{Garrison2015}
Garrison KA, Scheinost D, Finn ES, Shen X, Constable RT (2015) The
  (in)stability of functional brain network measures across thresholds.
  {\em NeuroImage, S1053-8119}(15):00,428--0

\bibitem[{Gazzaniga(2009)}]{Gazzaniga2013b}
Gazzaniga MS (ed)  (2009) The Cognitive Neurosciences. MIT Press

\bibitem[{Ghrist(2014)}]{ghrist2014elementary}
Ghrist R (2014) Elementary applied topology, 1st edition. Createspace

\bibitem[{Ginestet et~al(2011)Ginestet, Nichols, Bullmore, and
  Simmons}]{Ginestet2011}
Ginestet CE, Nichols TE, Bullmore ET, Simmons A (2011) Brain network analysis:
  separating cost from topology using cost-integration. {\em PLoS ONE, 6}(7):e21,570

\bibitem[{Giusti et~al(2015)Giusti, Pastalkova, Curto, and
  Itskov}]{giusti2015clique}
Giusti C, Pastalkova E, Curto C, Itskov V (2015) Clique topology reveals
  intrinsic geometric structure in neural correlations. {\em Proceedings of the National Academy of the Sciences of the United States of America,
  112}(44):13,455--13,460, 

\bibitem[{Grosenick et~al(2015)Grosenick, Marshel, and
  Deisseroth}]{Grosenick2015}
Grosenick L, Marshel JH, Deisseroth K (2015) Closed-loop and activity-guided
  optogenetic control. {\em Neuron, 86}(1):106--139

\bibitem[{Khalid et~al(2014)Khalid, Kim, Chung, Ye, and
  Jeon}]{khalid2014tracing}
Khalid A, Kim BS, Chung MK, Ye JC, Jeon D (2014) Tracing the evolution of
  multi-scale functional networks in a mouse model of depression using
  persistent brain network homology. {\em NeuroImage, 101}:351--363

\bibitem[{Kim et~al(2014)Kim, Kang, Lee, Lee, Suh, Song, Oh, and
  Lee}]{kim2014morphological}
Kim E, Kang H, Lee H, Lee HJ, Suh MW, Song JJ, Oh SH, Lee DS (2014)
  Morphological brain network assessed using graph theory and network
  filtration in deaf adults. {\em Hearing Research, 315}:88--98

\bibitem[{Kozlov(2007)}]{kozlov2007combinatorial}
Kozlov D (2007) Combinatorial Algebraic Topology, vol~21. Springer Science \&
  Business Media

\bibitem[{Krug et~al(2015)Krug, Salzman, and Waddell}]{Krug2015}
Krug K, Salzman CD, Waddell S (2015) Understanding the brain by controlling
  neural activity. {\em Philosophical Transactions of the Royal Society or London B: Biological Sciences, 370}(1677):20140,201

\bibitem[{Langer et~al(2013)Langer, Pedroni, and J\"{a}ncke}]{Langer2013}
Langer N, Pedroni A, J\"{a}ncke L (2013) The problem of thresholding in
  small-world network analysis. {\em PLoS ONE, 8}(1):e53,199

\bibitem[{Lee et~al(2011)Lee, Chung, Kang, Kim, and
  Lee}]{lee2011discriminative}
Lee H, Chung MK, Kang H, Kim BN, Lee DS (2011) Discriminative persistent
  homology of brain networks. {\em In: Biomedical Imaging: From Nano to Macro, 2011
  IEEE International Symposium on}, IEEE, pp 841--844

\bibitem[{Lum et~al(2013)Lum, Singh, Lehman, Ishkanov, Vejdemo-Johansson,
  Alagappan, Carlsson, and Carlsson}]{lum2013extracting}
Lum P, Singh G, Lehman A, Ishkanov T, Vejdemo-Johansson M, Alagappan M,
  Carlsson J, Carlsson G (2013) Extracting insights from the shape of complex
  data using topology. {\em Scientific Reports, 3}

\bibitem[{Medaglia et~al(2015)Medaglia, Lynall, and Bassett}]{Medaglia2015}
Medaglia JD, Lynall ME, Bassett DS (2015) Cognitive network neuroscience. {\em Journal of
  Cognitive Neuroscience, 27}(8):1471--1491

\bibitem[{Nanda and Sazdanovi{\'c}(2014)}]{nanda2014simplicial}
Nanda V, Sazdanovi{\'c} R (2014) Simplicial models and topological inference in
  biological systems. {\em In: Discrete and Topological Models in Molecular Biology},
  Springer, pp 109--141

\bibitem[{Nicolau et~al(2011)Nicolau, Levine, and
  Carlsson}]{nicolau2011topology}
Nicolau M, Levine AJ, Carlsson G (2011) Topology based data analysis identifies
  a subgroup of breast cancers with a unique mutational profile and excellent
  survival. {\em Proceedings of the National Academy of Sciences of the United States of America, 108}(17):7265--7270

\bibitem[{Petri et~al(2014)Petri, Expert, Turkheimer, Carhart-Harris, Nutt,
  Hellyer, and Vaccarino}]{petri2014homological}
Petri G, Expert P, Turkheimer F, Carhart-Harris R, Nutt D, Hellyer P, Vaccarino
  F (2014) Homological scaffolds of brain functional networks. {\em Journal of the Royal Society Interface,
  11}(101):20140,873

\bibitem[{Pirino et~al(2014)Pirino, Riccomagno, Martinoia, and
  Massobrio}]{pirino2014topological}
Pirino V, Riccomagno E, Martinoia S, Massobrio P (2014) A topological study of
  repetitive co-activation networks in in vitro cortical assemblies. {\em Physical Biology,
  12}(1):016,007--016,007

\bibitem[{Porter et~al(2009)Porter, Onnela, and Mucha}]{Porter2009}
Porter MA, Onnela JP, Mucha PJ (2009) Communities in networks. {\em Notices of the American Mathematical Society, 56}(9):1082--1097, 1164--1166

\bibitem[{Raj et~al(2012)Raj, Kuceyeski, and Weiner}]{Raj2012}
Raj A, Kuceyeski A, Weiner M (2012) A network diffusion model of disease
  progression in dementia. {\em Neuron, 73}(6):1204--1215

\bibitem[{Rubinov and Bassett(2011)}]{Rubinov2011}
Rubinov M, Bassett DS (2011) Emerging evidence of connectomic abnormalities in
  schizophrenia. {\em Journal of Neuroscience, 31}(17):6263--5

\bibitem[{Sala et~al(2014)Sala, Quatto, Valsasina, Agosta, and
  Filippi}]{Sala2014}
Sala S, Quatto P, Valsasina P, Agosta F, Filippi M (2014) {pFDR} and {pFNR}
  estimation for brain networks construction. {\em Statistics in Medicine, 33}(1):158--169

\bibitem[{Singh et~al(2008)Singh, Memoli, Ishkhanov, Sapiro, Carlsson, and
  Ringach}]{singh2008topological}
Singh G, Memoli F, Ishkhanov T, Sapiro G, Carlsson G, Ringach DL (2008)
  Topological analysis of population activity in visual cortex. {\em Journal of Vision, 8}(8):11

\bibitem[{Sporns(2014)}]{Sporns2014}
Sporns O (2014) Contributions and challenges for network models in cognitive
  neuroscience. {\em Nature Neuroscience, 17}(5):652--660

\bibitem[{Stam(2014)}]{Stam2014}
Stam CJ (2014) Modern network science of neurological disorders. {\em Nature Reviews
  Neuroscience, 15}(10):683--695

\bibitem[{Stolz(2014)}]{stolz2014computational}
Stolz B (2014) Computational topology in neuroscience. Master's thesis,
  University of Oxford

\bibitem[{Sulzer et~al(2013)Sulzer, Haller, Scharnowski, Weiskopf, Birbaumer,
  Blefari, Bruehl, Cohen, DeCharms, Gassert, Goebel, Herwig, LaConte, Linden,
  Luft, Seifritz, and Sitaram}]{Sulzer2013}
Sulzer J, Haller S, Scharnowski F, Weiskopf N, Birbaumer N, Blefari ML, Bruehl
  AB, Cohen LG, DeCharms RC, Gassert R, Goebel R, Herwig U, LaConte S, Linden
  D, Luft A, Seifritz E, Sitaram R (2013) Real-time {fMRI} neurofeedback:
  progress and challenges. {\em NeuroImage, 76}:386--399

\bibitem[{Szatmary and Izhikevich(2010)}]{Szatmary2010}
Szatmary B, Izhikevich EM (2010) Spike-timing theory of working memory. {\em PLoS
  Computational Biology, 6}(8)

\bibitem[{Taylor et~al(2015)Taylor, Klimm, Harrington, Kram{\'a}r, Mischaikow,
  Porter, and Mucha}]{taylor2015topological}
Taylor D, Klimm F, Harrington HA, Kram{\'a}r M, Mischaikow K, Porter MA, Mucha
  PJ (2015) Topological data analysis of contagion maps for examining spreading
  processes on networks. {\em Nature Communications, 6}

\bibitem[{Xia et~al(2015)Xia, Feng, Tong, and Wei}]{xia2015persistent}
Xia K, Feng X, Tong Y, Wei GW (2015) Persistent homology for the quantitative
  prediction of fullerene stability. {\em Journal of Computational Chemistry,
  36}(6):408--422

\bibitem[{Zhou et~al(2012)Zhou, Gennatas, Kramer, Miller, and
  Seeley}]{Zhou2012}
Zhou J, Gennatas ED, Kramer JH, Miller BL, Seeley WW (2012) Predicting regional
  neurodegeneration from the healthy brain functional connectome. {\em Neuron,
  73}(6):1216--1227

\end{thebibliography}

\end{document}